\documentclass[conference]{IEEEtran}
\usepackage{amsmath,amsthm}
\usepackage{cite}
\usepackage{graphicx}
\usepackage{epstopdf}
\usepackage{amsfonts,amsmath,amssymb}
\usepackage{cite}
\usepackage{graphicx}
\usepackage{url}
\usepackage{bm}
\usepackage{bbm}
\usepackage{amssymb}

\begin{document}

\title{{Multimode Entangled States in the Lossy Channel}}
\author{
\IEEEauthorblockN{Nedasadat Hosseinidehaj and Robert Malaney}
\IEEEauthorblockA{School of Electrical Engineering  \& Telecommunications,\\
The University of New South Wales,\\
Sydney, NSW 2052, Australia.\\
neda.hosseini@unsw.edu.au, r.malaney@unsw.edu.au}
}
\vspace{-1cm}
\maketitle
\begin{abstract}
In this work we analyse the structure of highly-entangled multimode  squeezed states, such as those  generated by broadband pulses undergoing type-II parametric down-conversion (PDC).  Such down-conversion has previously been touted as a natural and  efficient means of cluster-state generation, and therefore a viable future pathway to quantum computation. We first detail how broadband PDC processes lead directly to a series of orthogonal supermodes that are linear combinations of the original frequency modes. We then calculate the total squeezing of the multimode entangled states when they are assumed to be measured by an ideal homodyne detection in which all  supermodes of the  states are detected by an optimally shaped local oscillator (LO) pulse. For comparison, squeezing of the same entangled states are calculated when  measured by a lower-complexity homodyne detection scheme that exploits an unshaped LO pulse. Such calculations illustrate the cost, in the context of squeezing, of moving from higher complexity (harder to implement)  homodyne detection  to lower-complexity (easier-to-implement) homodyne detection. Finally, by studying the degradation in squeezing of the supermodes under photonic loss,  multimode entangled state evolution through an attenuation channel is determined. The results reported here push us towards a fuller understanding of the real-world transfer of cluster-states when they take the form of highly-entangled multimode states in frequency space.

\end{abstract}

\section{Introduction}

Numerous photonic architectures have been proposed for the implementation of quantum information processing, with measurement-based   linear optical computing perhaps being the most well known \cite{kk1}. However, practical scalability issues continue to hinder progress in this well-known architecture. A quite different architecture for quantum information processing that has garnered interest in recent times is the use of  measurements of single nodes contained within  a highly entangled multimode (multipartite) state - that takes the form of a cluster state, e.g. \cite {kk9}. A key problem for this latter paradigm  is the generation of the cluster state itself - a difficult task due to the intrinsic large-scale entanglement required. A potentially promising avenue to solve this problem is the use of frequency combs as input to the down-conversion process in an optical parameter oscillator,  e.g. \cite{kk14, kk15, kk16, kkx}.

In this work we will not investigate specific cluster state generation techniques, but rather consider the evolution of such states through photonic-loss channels when these states are indeed in the form of highly-entangled multimode states. We will be particularly interested in the total entanglement that resides within such states following their transmission through a photonic-loss channel. Since we will investigate the entanglement properties as measured by  homodyne techniques, our journey will also highlight the importance of proper pulse shaping of the LO in any quantum protocol that is based on down converted broadband pulses followed by homodyne detection (e.g. see \cite{Neda1} for one example of such a protocol). Our work therefore, not only probes the transfer (communication through a quantum channel) of specifically engineered cluster states in their own right, but also the transfer and detection of generic multimode states produced by broadband laser pulses undergoing a PDC process.

In continuous-variable (CV) domain, type-II PDC is known as a source of entangled squeezed quantum states. In type-II PDC,  a photon of the incoming pump beam spontaneously decays (in a non-linear crystal) into a pair of photons, known as the signal and the idler. The signal and the idler are in orthogonal polarizations,  forming an entangled squeezed state \cite{1987, 1997, 1997-2}. Standard type-II PDC sources do not generate a single entangled squeezed state, but (simultaneously) a multitude of entangled squeezed states in frequency modes  \cite{example-2000, 2001, silber2009, christ2009}. In general, each frequency mode of the signal beam can be correlated with all frequency modes of the idler beam. However, in entanglement-based quantum communication protocols, such as quantum key distribution (QKD) and quantum teleportation where the PDC state (i.e, the output of type-II PDC) is mostly used as the entanglement resource, the multimode structure of such a state is usually ignored.

In order to analyse the multimode structure of the PDC state a new basis is defined in which two sets of orthogonal (frequency-decorrelated) broadband spectral modes, called \emph{supermodes} are introduced (one set for the signal beam and the other set for the idler beam). These supermodes are linear combinations of the original, single frequency modes \cite{example-2000, 2001, silber2011, christ2011, christ2012,christ2013}. In such a modal representation, each supermode of the signal beam is correlated with only one supermode of the idler beam. Thus, instead of describing the PDC state as an entangled squeezed state in the frequency mode basis, the PDC state can be described as a set of independent entangled squeezed states in the supermode basis. Such a supermode basis can also be defined for type-I PDC of femtosecond-frequency combs \cite{kkx, PDCI-2006, PDCI-2007, PDCI-2010, comb2014}. 

The remainder of the paper is as follows: In Section II the  structure of the PDC state is presented and homodyne detection is discussed.  Section III presents our simulation results for different LO pulses, and Section IV presents the effects of photonic loss.


\section{System Model}
\subsection{Multimode Squeezed Entangled States}

Homodyne detection is widely used in quantum communication protocols to measure the quadrature statistics of a quantum state, e.g.
 \cite{homodyne2000, homodyne2007, homodyne2013}.
 In many homodyne detections of the PDC state, an unshaped LO pulse (which is actually the pump laser pulse) is utilized for the quadrature measurements of the signal and idler beams.
As a result, the quadrature statistics of one mode of the signal and one mode of the idler beam  are measured. From these quadrature statistics the correlation between these two  modes can be obtained. In some cases this may only be  a fraction of the total correlations between the signal and idler beams. However, the total correlations of the PDC state can be obtained by capturing all the orthogonal supermodes of the state in the homodyne detection which can be realized by shaping the LO pulse in the spectral form of the orthogonal supermodes \cite{Opatrny}.

In type-II PDC, a pump beam with spectral amplitude $\alpha \left( \omega  \right) $ is first frequency doubled. 
The nonlinear crystal is then pumped with the frequency-doubled pump beam where a photon of the incoming beam spontaneously decays into an orthogonally-polarized photon pair (the signal and the idler), forming an entangled squeezed state. This process can be described by the Hamiltonian \cite{1987, 1997, 1997-2, example-2000, 2001, Ham2008},
\begin{equation}\label{H}
{{\hat H}_{PDC}} = \zeta \int {\int {d{\omega _s}} } d{\omega _i}f\left( {{\omega _s},{\omega _i}} \right){{\hat a}^\dag }\left( {{\omega _s}} \right){{\hat b}^\dag }\left( {{\omega _i}} \right) + h.c.,
\end{equation}
where $\hat a^\dag \left( {{\omega _s}} \right)$ is the photon creation operator associated with a signal mode of frequency ${{\omega _s}}$, $\hat b^\dag \left( {{\omega _i}} \right)$ is the photon creation operator associated with an idler mode of frequency ${{\omega _i}}$, and $\zeta $ denotes the overall efficiency of the PDC process. Note \emph{h.c.} is the Hermitian conjugate. The function $f\left( {{\omega _s},{\omega _i}} \right)$ is the joint spectral amplitude of the emitted photon pairs, which indicates the coupling strength between modes at frequencies ${{\omega _s}}$ and ${{\omega _i}}$, and is given by
\begin{equation}\label{f}
f\left( {{\omega _s},{\omega _i}} \right) = \frac{1}{{\sqrt N }}\alpha '\left( {\omega '} \right)\Phi \left( {{\omega _s},{\omega _i}} \right),
\end{equation}
where $\alpha '\left( {\omega '} \right)$ is the spectral amplitude of the frequency-doubled pump beam at frequency ${\omega '} = {\omega _s} + {\omega _i}$, and $\Phi \left( {{\omega _s},{\omega _i}} \right)$ is the phase-matching function. The function $f\left( {{\omega _s},{\omega _i}} \right)$ is normalized via $N$ such that $\int {\int {d{\omega _s}} } d{\omega _i}{\left| {f\left( {{\omega _s},{\omega _i}} \right)} \right|^2} = 1$. Note we assume a real-valued function for $f\left( {{\omega _s},{\omega _i}} \right)$. For more general functions (e.g. such as used in \cite{2001}) details of our calculations will change, but the broad results will remain.

The PDC state can be described as $\left| {PDC} \right\rangle  = {{\hat U}_{PDC}}\left| 0 \right\rangle \left| 0 \right\rangle $, where the unitary operator ${{\hat U}_{PDC}}$ is given by ${{\hat U}_{PDC}} = \exp \left( { - \frac{i}{\hbar }{{\hat H}_{PDC}}} \right)$. Such a representation of the PDC state is not easy to analyse, as there are infinitely many frequency-correlated modes in the signal and idler beams.

By performing a singular-value decomposition (SVD) on the function $f\left( {{\omega _s},{\omega _i}} \right)$, we are able to express it in a frequency-decorrelated modal representation \cite{example-2000, 2001, silber2011, christ2011, christ2012,christ2013}, i.e.
\begin{equation}\label{svd}
f\left( {{\omega _s},{\omega _i}} \right) = \sum\limits_k {{c_k}} {\psi _k}\left( {{\omega _s}} \right){\varphi _k}\left( {{\omega _i}} \right),
\end{equation}
where $\left\{ {{\psi _k}\left( {{\omega _s}} \right)} \right\}$ and $\left\{ {{\varphi _k}\left( {{\omega _i}} \right)} \right\}$ are two (real-valued) orthonormal basis sets. Each pair of functions ${{\psi _k}\left( {{\omega _s}} \right)}$ and ${{\varphi _k}\left( {{\omega _i}} \right)}$ describe a \emph{supermode} with the amplitude $c_k$, such that ${\sum\limits_k {\left| {{c_k}} \right|} ^2} = 1$.
The type-II PDC Hamiltonian can now be expressed in terms of supermodes, as
\begin{equation}\label{H-super}
{{\hat H}_{PDC}} = \zeta\sum\limits_k {{c_k}} \left( {\hat A_k^\dag \hat B_k^\dag  + {{\hat A}_k}{{\hat B}_k}} \right),
\end{equation}
where ${\hat A_k^\dag }$ (${\hat B_k^\dag }$) is the photon creation operator associated with a signal (idler) supermode, defined as
\begin{equation}\label{operator-super}
\begin{array}{l}
{{\hat A}_k} = \int {d{\omega _s}{\psi _k}\left( {{\omega _s}} \right)} \hat a\left( {{\omega _s}} \right),\\
\\
{{\hat B}_k} = \int {d{\omega _i}{\varphi _k}\left( {{\omega _i}} \right)} \hat b\left( {{\omega _i}} \right).
\end{array}
\end{equation}
Due to the orthonormal property of the supermode functions, i.e., $\left\{ {{\psi _k}\left( {{\omega _s}} \right)} \right\}$ and $\left\{ {{\varphi _k}\left( {{\omega _i}} \right)} \right\}$, the supermode field operators satisfy the commutation relations, $\left[ {{{\hat A}_i},\hat A_j^\dag } \right] = \left[ {{{\hat B}_i},\hat B_j^\dag } \right] = {\delta _{ij}}$.

Such a representation shows that the type-II PDC generates an ensemble of independent twin-beam supermode squeezed states, for which ${r_k} = \zeta{c_k}$ determines the twin-beam squeezing in supermode $k$. Note, in terms of dB the twin-beam squeezing in supermode $k$ is defined by $ - 10{\log _{10}}\left( {\exp ( - 2{r_k})} \right)$. The quadrature operators associated with a signal supermode ($\hat Q_k^a,\,\hat P_k^a$), and an idler supermode ($\hat Q_k^b,\,\hat P_k^b$) can now be defined as
\begin{equation}\label{quadrature-super}
\begin{array}{l}
\hat Q_k^a = {{\hat A}_k} + \hat A_k^\dag ,\,\hat P_k^a = i\left( {\hat A_k^\dag  - {{\hat A}_k}} \right),\\
\\
\hat Q_k^b = {{\hat B}_k} + \hat B_k^\dag ,\,\hat P_k^b = i\left( {\hat B_k^\dag  - {{\hat B}_k}} \right).
\end{array}
\end{equation}
The vector of the supermode quadrature operators for the PDC state with $n$ supermodes can be defined as $\hat R = \left( {\hat Q_1^a,\,\hat P_1^a,\hat Q_1^b,\,\hat P_1^b, \ldots ,\hat Q_n^a,\,\hat P_n^a,\hat Q_n^b,\,\hat P_n^b} \right)$. The covariance matrix (CM) of the PDC state in terms of the supermode quadrature operators is defined as ${M_{ij}} = \frac{1}{2}\left\langle {{{\hat R}_i}{{\hat R}_j} + {{\hat R}_j}{{\hat R}_i}} \right\rangle  - \left\langle {{{\hat R}_i}} \right\rangle \left\langle {{{\hat R}_j}} \right\rangle $, where $\left\langle . \right\rangle $ denotes the first moment of the supermode quadrature operator. Each twin-beam supermode squeezed state with squeezing $r_k$  is a Gaussian state \cite{kk9}, and can be described by a CM in the following form
\begin{equation}\label{CM-super}
{M_k} = \left( {\begin{array}{*{20}{c}}
{\cosh (2{r_k})I}&{\sinh (2{r_k})Z}\\
{\sinh (2{r_k})Z}&{\cosh (2{r_k})I}
\end{array}} \right),
\end{equation}
where $I$ is a $2 \times 2$ identity matrix and $Z = diag\left( {1, - 1} \right)$. For a pure PDC state the total squeezing is given by ${r_{tot}} = \sum\limits_k {{r_k}} $, and the total logarithmic negativity (as a measure of entanglement) is given by $E_{tot} = \sum\limits_k { - {{\log }_2}} \left( {\exp \left( { - 2{r_k}} \right)} \right)$ \cite{kk9}.


\subsection{Homodyne Detection}

To measure the quadrature statistics of an optical pulse through a homodyne detector, the signal pulse is combined in a 50:50 beam splitter with a strong LO pulse. In general, the signal pulse consists of a continuum of frequency modes with field operators $\hat a\left( \omega  \right)$ and ${{\hat a}^\dag }\left( {{\omega }} \right)$. Let us assume that the LO pulse can be described by a normalized real-valued spectral function $g(\omega )$. Through the measurement of the difference-photocurrent statistics, the signal pulse quadrature operator, $\hat X =  \left( {\hat A\exp \left( { - i\varphi } \right) + {{\hat A}^\dag }\exp \left( {i\varphi } \right)} \right)$ is measured, where $\hat A = \int {d\omega } g(\omega )\hat a\left( \omega  \right)$, and where $\varphi$ is the phase of the LO pulse. Note, for $\varphi=0$, the quadrature operator $\hat Q = \left( {\hat A + {{\hat A}^\dag }} \right)$ and for $\varphi=\pi /2$, the quadrature operator $\hat P = i\left( {{{\hat A}^\dag } - \hat A} \right)$ of the signal pulse can be measured. Thus, an LO pulse can be used to `capture' a supermode (with field operator $\hat A$) in the signal pulse, and measure the associated quadrature operators \cite{homodyne2000}.

The above discussion informs us that if the LO pulse in the homodyne detection of the signal (idler) beam of the PDC state is shaped so that it is in the spectral form of the supermode function ${\psi _k}\left( {{\omega _s}} \right)$ (${\varphi _k}\left( {{\omega _i}} \right)$), one can measure the quadrature operators of the corresponding signal (idler) supermode given by Eq.~\eqref{quadrature-super}. In this case according to Eq.~(\ref{svd}) the corresponding twin-beam supermode squeezing is given by ${r_k} = \zeta \int {\int {d{\omega _s}} } d{\omega _i}{\psi _k}\left( {{\omega _s}} \right)f\left( {{\omega _s},{\omega _i}} \right){\varphi _k}\left( {{\omega _i}} \right)$.


\section{Optimal and Sub-optimal Detection}

Here, we consider a typical (experimental) type-II PDC process, where the spectral amplitude of the frequency-doubled pump beam  $\alpha '\left( {\omega '} \right)$, and the phase-matching function $\Phi \left( {{\omega _s},{\omega _i}} \right)$ are given as
\begin{equation}\label{pump-phasematch}
\begin{array}{l}
\alpha '\left( {\omega '} \right) = \alpha '\left( {{\omega _s} + {\omega _i}} \right) = \exp \left( { - \frac{{{{\left( {{\omega _s} + {\omega _i} - 2{\omega _p}} \right)}^2}}}{{2\sigma _p^2}}} \right),\,\,\\
\\
\Phi \left( {{\omega _s},{\omega _i}} \right) = \sin c\left( { - \frac{{{k_s}\left( {{\omega _s} - {\omega _p}} \right) + {k_i}\left( {{\omega _i} - {\omega _p}} \right)}}{2}} \right),
\end{array}
\end{equation}
where ${\sigma _p}$ is the pump bandwidth and $\omega _p$ is the central frequency of the original pump beam. In the phase-matching function we have ${k_s} = L({{k'_p}} - {{k'_s}})$ and ${k_i} = L({{k'_p}} - {{k'_i}})$ where $L$ is the crystal length, and where $k'_p$, $k'_s$, and $k'_i$ are the inverse of group velocities at the frequencies $2{\omega _p}$, ${\omega _p}$, and ${\omega _p}$, respectively. Unless otherwise stated, here we consider the values of a typical experiment in \cite{example-2000}, where $k_s=0.061$ ps and $k_i=0.213$ ps corresponding to a value of $L=0.8$mm.



We assume the PDC source to be pumped by a laser source delivering ultrafast optical pulses with ∼6 nm FWHM (∼140 fs) centered at 795 nm with a repetition rate of 76 MHz \cite{comb2014, comb2015}. For such a pulse the frequency bandwidth is approximately $2.8480 \times {10^{12}}$ Hz FWHM.


We approximate the function $f\left( {{\omega _s},{\omega _i}} \right)$ by subdividing the frequency range $[0,2w_p+3\sigma_p]$ (for both $\omega_s$ and $\omega_i$) into $n=1000$ discrete bins which allows us to form an $n \times n$ matrix.
Performing an SVD on this $n \times n$ matrix (i.e., Eq.~(\ref{svd})), we will have $n$ supermodes in both the signal and idler beams. The $k$-th supermode of the signal beam is only correlated with the $k$-th supermode of the idler beam, forming a twin-beam supermode squeezed state with a twin-beam squeezing $r_k$ (calculated from the SVD). Then, we can calculate ${r_{tot}} = \sum\nolimits_{k = 1}^n {{r_k}} $, which is a good approximation of the total squeezing of the PDC state. Note also, we assume unit efficiency for the PDC process, i.e., $\zeta=1$.
In Fig.~1, we plot $r_k$ of the first 20 supermodes (out of $n=1000$ supermodes) for different values of $\sigma_p$. According to Fig.~1, squeezing amplitudes $r_k$  exponentially (approximately) decay for higher-order supermodes (i.e., higher values of $k$), with the decay rates decreasing with decreasing  pump bandwidth. Note, for ${\sigma _p={\rm{1}}{\rm{.4240}} \times {10^{12}}}$ Hz, the squeezing included in the leading 20 supermodes is approximately $50\% $ of the total squeezing $r_{tot}=6$.


\begin{figure}[!t]
    \begin{center}
   {\includegraphics[width=3.3 in, height=2 in]{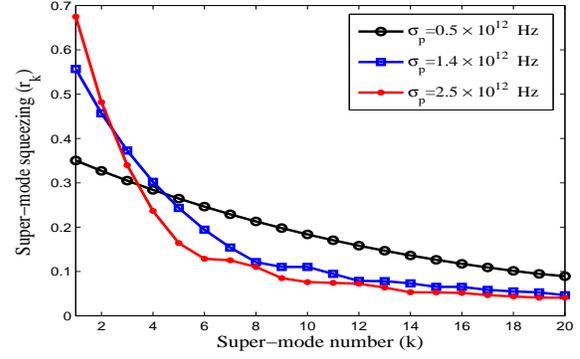}}
    \caption{The twin-beam squeezing of the first 20 supermodes.}\label{fig1}
    \end{center}
\end{figure}

We also plot the total squeezing $r_{tot}$ of the PDC state as a function of the pump bandwidth ${\sigma _p}$ in Fig.~2 (top figure). As it can be seen the total squeezing can be increased by decreasing the pump bandwidth. Such a squeezing improvement can be explained because a narrower bandwidth (i.e., smaller ${\sigma _p}$) leads to a larger value of $\alpha '\left( {\omega '} \right)$, resulting in stronger coupling between frequency modes of the signal and idler beams in the function $f\left( {{\omega _s},{\omega _i}} \right)$. In all our calculations thus far, the total squeezing $r_{tot}$ is calculated assuming the LO pulses are shaped in the form of the supermode functions (optimal detection).


\begin{figure}[!t]
    \begin{center}
   {\includegraphics[width=3.3 in, height=4.5 in]{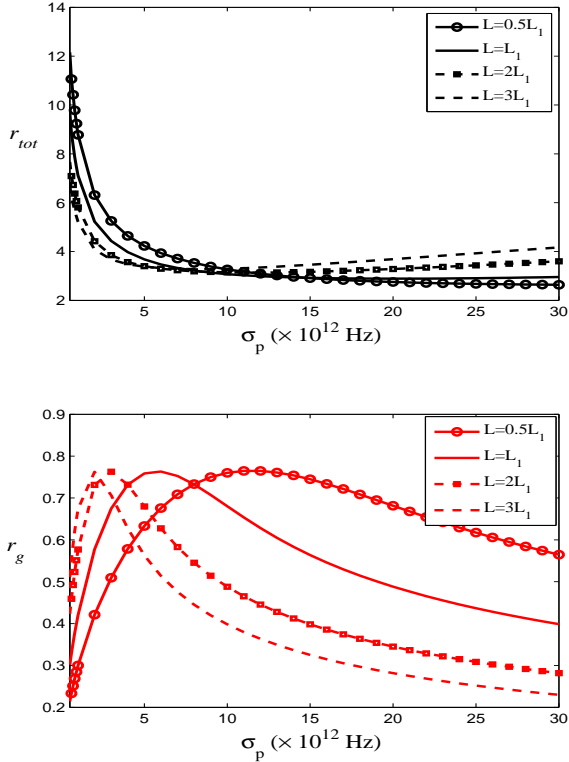}}
    \caption{The total squeezing $r_{tot}$ of the PDC state as a function of the pump bandwidth, $\sigma_p$ (top figure). The squeezing $r_g$ of the PDC state when it is homodyne-detected by  unshaped pulses as a function of $\sigma_p$ (bottom figure).  Note, the plots show the squeezing for different crystal lengths. Here $L_1=0.8$mm, and the solid unmarked line ($L=L_1$)   leads to values $k_s=0.061$ ps and $k_i=0.213$ ps.}\label{fig2}
    \end{center}
\end{figure}

Now, we consider homodyne detection of the PDC state without shaping LO pulses (in the form of $\left\{ {{\psi _k}\left( {{\omega _s}} \right)} \right\}$ and $\left\{ {{\varphi _k}\left( {{\omega _i}} \right)} \right\}$). In fact, the LO pulse for the homodyne detection of each signal and idler beams is directly taken from the laser source, which means the LO pulse is actually in the form of pump spectral function, i.e., $\alpha \left( \omega  \right)$. Thus, we consider the normalized functions

\begin{equation}\label{unshaped-r}
{g_s}\left( \omega  \right) = {g_i}\left( \omega  \right) = \frac{1}{{\sqrt {{N_p}} }}\exp \left( { - \frac{{{{\left( {\omega  - {\omega _p}} \right)}^2}}}{{2\sigma _p^2}}} \right)
\end{equation}
as the two unshaped LO pulses, where ${g_s}\left( \omega  \right)$ (${g_i}\left( \omega  \right)$) is the LO pulse for the homodyne detection of the signal (idler) beam, and $N_p$ is the normalization constant.


As discussed earlier, in the homodyne detection an LO pulse with the function ${g_s}\left( \omega  \right)$ (${g_i}\left( \omega  \right)$) captures a supermode in the signal (idler) beam, and the associated quadrature operator can be measured. The squeezing associated with the captured supermodes can be given by
\begin{equation}\label{unshaped-r}
{r_g} = \zeta \int {\int {d{\omega _s}} } d{\omega _i}{g_s}\left( {{\omega _s}} \right)f\left( {{\omega _s},{\omega _i}} \right){g_i}\left( {{\omega _i}} \right).
\end{equation}

Here we consider the PDC state described by Eq.~(\ref{pump-phasematch}) which is now homodyne detected by the unshaped LO pulses ${g_s}\left( \omega  \right)$  and ${g_i}\left( \omega  \right)$. In Fig.~2 (bottom figure) we plot the obtained squeezing, $r_g$, by such a homodyne detection as a function of the pump bandwidth, $\sigma_p$.
As can be seen the squeezing $r_g$ is first increased by increasing the pump bandwidth, because a wider LO pulse is able to capture a wider supermode (i.e. a linear combination of more frequency modes). However, the correlation between these frequency modes is decreased by increasing the pump bandwidth. Thus, there is a peak in the value of $r_g$, which means there is an optimal value of the pump bandwidth to maximize the obtainable squeezing when the PDC state is homodyne-detected by unshaped LO pulses.

Comparison between the top and bottom plots of Fig.~2 clearly shows the  cost of the sub-optimal homodyne detection of the PDC state  relative to the optimal homodyne detection. That is, Fig.~2 shows that for all values of the pump bandwidth, the LO pulses need to be properly shaped in order to measure the total squeezing of the PDC state otherwise only a small fraction of the total squeezing can be measured. For instance, for $\sigma_p=8  \times {10^{12}}$ and $L=L_1$, $r_g$ is  approximately $20\% $ of the total squeezing $r_{tot}$. Note that, the parameters that are varied in these plots represent perhaps the two most fundamental variables under the experimentalist's control - namely the bandwidth of the pump and the length of the non-linear crystal. Opposing effects are at play here\footnote{Note when $\sigma_p$ increases, the coupling between frequency modes of the signal and idler beams becomes weaker, however, the number of frequency modes contributing to the total squeezing increases.}, but a particular observation worthy of note is the fact that as $L$ increases and the bandwidth decreases the mismatch between optimal and suboptimal detection can decrease.\footnote{Of course other processes can be utilized to reduce such a mismatch, e.g. frequency filtering at the sending station. However, such processes will usually remove the cluster-state configuration, the evolution of which is of prime interest to us here.}




Note, if we reduce the bandwidth of the unshaped LO pulses, $\sigma _p$, such that the LO pulses approach a delta function, i.e., ${g_s}\left( \omega  \right) = {g_i}\left( \omega  \right) = \delta \left( {\omega  - {\omega _p}} \right)$, the captured twin-beam supermode squeezed state approaches the twin-beam frequency-mode (the mode associated with the central frequency ${{\omega _p}}$) squeezed state.



\section{Lossy Channels}
In the previous section,
we assumed there is no loss in the PDC state. In this section we analyse the evolution of the PDC state through a lossy channel in terms of the transferred squeezing levels - a metric important for many quantum information protocols. We note  the transmission of the PDC state over a lossy channel has been previously analysed  in terms of the achievable quantum communication rates \cite{christ2012}.

We assume the idler beam of the PDC state is kept unchanged at the source, while the signal beam is transmitted through a lossy channel. For simplicity, we also assume the lossy channel is a fixed-attenuation channel with transmissivity $\tau$, and that losses are independent of the frequency (of the signal beam).
Since the channel loss is independent of the frequency, the mode functions $\left\{ {{\psi _k}\left( {{\omega _s}} \right)} \right\}$ and $\left\{ {{\varphi _k}\left( {{\omega _i}} \right)} \right\}$ of the PDC state remain unchanged \cite{filter}. Hence, all the twin-beam supermode squeezed states evolve independently from each other through the channel. Thus, the total logarithmic negativity of the PDC state at the output of the fixed-attenuation channel can be calculated through the evolution of the CM of each twin-beam supermode squeezed state. Each twin-beam supermode squeezed state with an initial CM as given by Eq.~\eqref{CM-super},  is after transmission of the signal beam (through a fixed-attenuation channel with transmissivity $\tau$) described by an evolved CM of the following form

\begin{equation}\label{CM-super_loss}
M_k^{loss} = \left( {\begin{array}{*{20}{c}}
{\cosh (2{r_k})I}&{\sqrt \tau  \sinh (2{r_k})Z}\\
{\sqrt \tau  \sinh (2{r_k})Z}&{\left( {\tau \cosh (2{r_k}) + 1 - \tau } \right)I}
\end{array}} \right).
\end{equation}
As such, the (approximated) total logarithmic negativity of the mixed PDC state is given by $E_{tot} = \sum\nolimits_{k = 1}^n {{E_k}} $, where ${E_k}$ is the logarithmic negativity of the evolved twin-beam supermode squeezed state described by the CM $M_k^{loss}$.

Note, given a Gaussian state with a CM $M = ( {A,C;{C^T},B} )$, where $A=A^T$, $B=B^T$, and $C$ are $2 \times 2$ real matrices, the logarithmic negativity is given by ${E}\left( {{M}} \right) = \max \left[ {0, - {{\log }_2}\left( {{\nu _ - }} \right)} \right]$, where ${\nu _ - }$ is the smallest symplectic eigenvalue of the partially transposed $M$. This eigenvalue is given by $
\nu _ - ^2 = \left( {\Delta  - \sqrt {{\Delta ^2} - 4\det M} } \right)/2
$, where $\Delta = \det A + \det B - 2\det C$ \cite{kk9}.

Considering the pure PDC state described by Eq.~(\ref{pump-phasematch}) as the initial PDC state, we plot in Fig.~3 the total logarithmic negativity, $E_{tot}$, of the mixed PDC state at the output of the channel as a function of the channel transmissivity $\tau$ and the pump bandwidth $\sigma_p$. As it can be seen, the total logarithmic negativity (similar to the total squeezing) of the PDC state is increased by decreasing the pump bandwidth, while it decreases with increasing channel loss. These results quantify, for a given input bandwidth, how much loss a channel can tolerate in order to achieve some target total entanglement. As such, they should prove useful in designing specific protocols which utilize optimal detection of the embedded entanglement within the quantum state.

\begin{figure}[!t]
    \begin{center}
   {\includegraphics[width=3.3 in, height=2.4 in]{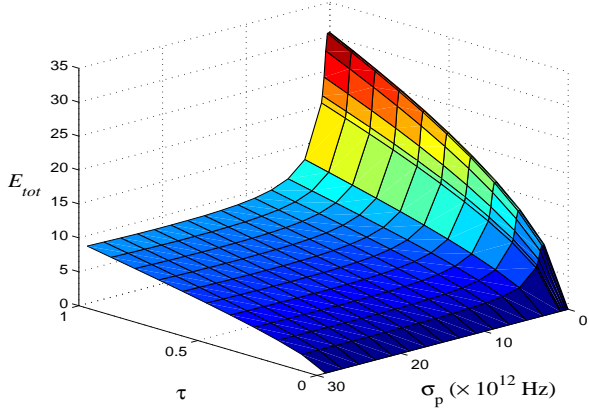}}
    \caption{The total logarithmic negativity, $E_{tot}$, of the mixed PDC state as a function of the channel transmissivity $\tau$ and pump bandwidth $\sigma_p$.}\label{fig4}
    \end{center}
\end{figure}

\section{conclusion}

In this work we have explored the importance of optimal pulse shaping in the homodyne measurements of any cluster state which is embedded in a multipartite entangled quantum state. Our results illustrate the trade off in complexity (optimal pulse shaping) versus quality (total entanglement) that arises in any information processing using such states. We have also determined how the total entanglement of the cluster state (as determined via optimal pulse shaping) is impacted by photon loss in a fixed attenuation channel. Future work could  include variable attenuation channels, and the impact of non-Gaussian operations such as photon subtractions at the sending station. Determination of the range  over which significant supermode-multiplexing gain is viable once frequency-dependent loss is considered would also be useful. Frequency-dependent losses will, in part, determine the transceiver dimensions needed  to coherently transfer cluster states over a specified distance.


\end{document}